\documentclass[reprint,amsmath,amssymb,aps,pre]{revtex4-1}

\usepackage{graphicx,amssymb}

\begin{document}

%
%
%

\title{Patterns, localized structures and fronts in a reduced model of
clonal plant growth}

\author{Daniel Ruiz-Reyn\'es$^1$, Luis Mart\'{\i}n$^1$, Emilio Hern\'andez-Garc\'ia$^1$,
Edgar Knobloch$^2$, Dami\`{a} Gomila$^1$}
\affiliation{$^1$IFISC (CSIC-UIB), Campus Universitat de les
Illes Balears, 07122, Palma de Mallorca, Spain \\
$^2$Department of Physics, University of California at
Berkeley, Berkeley, CA 94720, USA}

\begin{abstract}
A simplified model of clonal plant growth is formulated,
motivated by observations of spatial structures in {\it
Posidonia oceanica} meadows in the Mediterranean Sea. Two
levels of approximation are considered for the scale-dependent
feedback terms. Both take into account mortality and clonal, or
vegetative, growth as well as competition and facilitation, but
the first version is nonlocal in space while the second is
local. Study of the two versions of the model in the
one-dimensional case reveals that both cases exhibit
qualitatively similar behavior (but quantitative differences)
and describe the competition between three spatially extended
states, the bare soil state, the populated state, and a pattern
state, and the associated spatially localized structures. The
latter are of two types, holes in the populated state and
vegetation patches on bare ground, and are organized within
distinct snaking bifurcation diagrams. Fronts between the three
extended states are studied and a transition between pushed and
pulled fronts identified. Numerical simulations in one spatial
dimension are used to determine front speeds and confront the
predictions from the marginal stability condition for pulled
fronts.
\end{abstract}

\date{August 29, 2020}

\maketitle

\section{Introduction}
\label{S1}

Vegetation distribution in space is often found to be spatially
inhomogeneous even in quite homogeneous terrains, a factor that
has been recognized as very relevant to understanding ecosystem
resilience, functioning and health
\cite{MeronBook,Rietkerk2004,Barbier2006,Watt1947pattern}.
Spatial self-organization of different types of plants has been
reported and modeled in a wide range of habitats, from arid or
semiarid environments to wetlands
\cite{Rietkerk2004,Barbier2006,Rietkerk2008,Meron2012} and,
more recently, in submerged seagrass meadows
\cite{ruiz2017fairy}. Detailed models of the competition of
plants for scarce water have been set up to understand pattern
formation in dry ecosystems
\cite{von2001diversity,Gilad2007mathematical,Gilad2007dynamics,Gandhi2018}.
In other approaches, water is not explicitly modeled but a
generic approach using an integral kernel, which takes into
account nonlocal competition processes with typical interaction
ranges, is used
\cite{lefever1997origin,MartinezGarcia2013,MartinezGarcia2014,FernandezOto14,Cisternas2020gapped}.
This last approach is more easily generalizable to situations
in which water is not the limiting resource driving competitive
interactions, as in the case of marine plants
\cite{ruiz2017fairy}.

In most of the previous models, propagation of vegetation over
a landscape is assumed to occur by seed dispersal, modeled as
an isotropic diffusion. Clonal growth by rhizome elongation,
however, has directional characteristics, a fact that has been
modeled at various levels of detail
\cite{ruiz2017fairy,ruiz2019angles,ruiz2019ABDaprox}. In
\cite{ruiz2017fairy} the so-called advection-branching-death
(ABD) model, describing the growth of clonal plants at a
landscape level, was introduced. This model consists of two
partial integro-differential equations for the evolution of the
density of shoots and apices of the plant. This model was
derived directly from the {\it microscopic} mechanisms involved
in clonal plant growth, namely rhizome elongation (which
appears in the model as an advection term), branching, and
death, and its parameters can be directly linked to rates and
quantities directly measured in underwater seagrass meadows
\cite{ruiz2017fairy}. Plant interactions were modeled in terms
of a nonlocal competition kernel. Although the model describes
vegetation distribution in two-dimensional space, the need to
include the direction of growth of the apices introduces a new
(angular) variable that makes this model effectively
three-dimensional and so carries a high computational cost.
Fortunately, this angular coordinate does not appear to play a
crucial role in most of the observed phenomenology
\cite{ruiz2019angles}. For this reason, in
\cite{ruiz2019ABDaprox} a single equation for the total density
of shoots that captures all the dynamical regimes of the ABD
model was proposed, the \emph{clonal-growth model}. Under
certain approximations this equation can be derived from the
full model, establishing a connection between the effective
parameters of the simple description and the biologically
relevant parameters of the full model.

Vegetation patterns can be \emph{extended}, in the sense that
the same type of biomass organization extends over a large area
(for example, homogeneous or periodic distributions) or
\emph{localized}. Examples of the later are a patch or several
patches of vegetation surrounded by bare soil, or a bare gap or
set of gaps on an otherwise vegetated area. Different extended
states can occur in different regions of a landscape, giving
rise to \emph{fronts} between them where they meet. Such fronts
can remain static or move, describing the invasion of one type
of biomass configuration by another or its retreat. In this
work we study in detail the spatially extended, localized, and
front solutions of two versions of the clonal-growth model
\cite{ruiz2019ABDaprox}, obtained by keeping the nonlocal
character of the interactions or replacing them by an effective
local description. In contrast to \cite{ruiz2019ABDaprox} we
restrict the present study to one spatial dimension, which
allows a more detailed analysis and comparison between the two
levels of approximation.

The models considered are described in Sec.~II.  The stationary
solutions of the resulting equations are presented and compared
in Sec.~III. The dynamics of fronts between different states,
both stable and unstable, are analyzed in Sec.~IV and in some
cases the predicted front speeds are compared with those
determined via direct numerical simulation.

\section{Model}
\label{S2}

In Ref.~\cite{ruiz2019ABDaprox} we proposed a sequence of
approximations that lead from the full ABD model to a single
differential equation describing the evolution of the shoot
density of a plant undergoing vegetative or clonal growth, the
clonal-growth model. In one spatial dimension, the model reads:
\begin{equation}
  \label{E1}
  \partial_t n = (\omega_b-\omega_{d}[n])n + d_0 \partial_{x}^2 n + d_1 n \partial_{x}^2 n + d_1 \left|\partial_x n\right|^2,
\end{equation}
where $n(x,t)$ is the plant shoot density, $\omega_b>0$ is the
branching rate of the plant, i.e., the birth term, and
$\omega_d[n]>0$ is the mortality rate, which depends on the
density. The terms involving derivatives with coefficients
$d_0$ and $d_1$ arise from rhizome elongation and branching in
the original model, and implement the peculiarities of clonal
growth. These coefficients may depend on environmental
conditions and hence on space, but we take both to be constant.
Note that the same coefficient $d_1$ appears in the last two
terms.

\subsection{Version I: full nonlocal interaction}

As a first level of description, hereafter \emph{version I},
we take Eq.~(\ref{E1}) and retain the original nonlocal
terms accounting for the interaction between plants
\cite{ruiz2017fairy}:
\begin{equation}
\label{E2}
\omega_d[n]= \omega_{d0} + \int \int
\mathcal{K}(\vec{r}-\vec{r}')(1-e^{-a n(\vec{r}')}) d\vec{r}' + b n^2.
\end{equation}
The first term $\omega_{d0}$ in (\ref{E2}) is the intrinsic
mortality (mortality in the absence of other plants). The
second term accounts for interactions across space in such a
way that the density of shoots at a given position can affect
the growth in a neighborhood weighted by the kernel
$\mathcal{K}(\vec{r}-\vec{r}')$. To better control the growth
of the solutions and prevent potential blow-up, we also
incorporate a local nonlinear term $bn^2$, where the parameter
$b$ determines the maximum value of the density even in cases
where nonlinear effects in the integral kernel might be
destabilizing. From the biological point of view this term
models local mechanisms of competition (at the level of single
shoots, for example) which are different from the non-local
ones modeled by the integral kernel.

The kernel $\mathcal{K}$ is taken to be the difference of two
normalized Gaussian functions $\mathcal{G}$, both with zero
mean but with different magnitudes ($\kappa$, $\mu > 0$) and
widths ($\sigma_{\kappa}$, $\sigma_{0}$):
\begin{equation}
\label{E3}
\mathcal{K}(\vec{r})=\kappa
\mathcal{G}(\sigma_\kappa,\vec{r})-\mu\mathcal{G}(\sigma_0,\vec{r}).
\end{equation}
The first term on the rhs of (\ref{E3}) accounts for all the
competitive effects, since it increases the mortality rate,
while the second accounts for facilitative effects.
Facilitation is taken here to reduce the external factors
increasing mortality only; the term $\omega_{d0}$, which is
independent of density, encodes these external effects. Hence,
the integrated facilitation needs to satisfy $\mu \leq
\omega_{d0}$ to ensure positive mortality and avoid unrealistic
creation of plants by this term. Here we choose $\mu =
\omega_{d0}$ for simplicity. The range of the competitive and
facilitative interaction is given by $\sigma_{\kappa}$ and
$\sigma_{0}$, respectively. We assume that $\sigma_k >
\sigma_0$, as appropriate for the observations in marine plants
\cite{ruiz2017fairy}.

\begin{figure}
    \centering
    \includegraphics[width=\columnwidth]{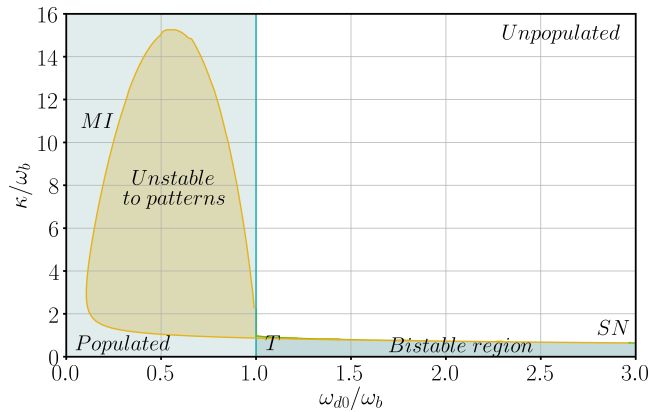}
    \caption{Phase diagram for version I of the model. The regions where the populated (P)
    and unpopulated (U) states are stable are shown in bright blue and white,
    respectively, while the coexistence region between these two states is
    shown in darker blue. The region where P is unstable to patterns is shown
    in yellow. The symbol $T$ refers to a transcritical bifurcation at
    $\omega_{d0}/\omega_b=1$, while $SN$ refers to a saddle-node bifurcation
    where P undergoes a fold. The curve labeled $MI$ corresponds to the onset
    of modulational (or Turing) instability that is responsible for the appearance
    of spatial patterns in the model. The parameters are
    $\omega_b = 0.06$ $ year^{-1}$, $b=1.25$ $ cm^4year^{-1}$, $\sigma_{\kappa}=$ $2851.4$ $cm$,
    $a=27.38$ $ cm^2$, $\sigma_0=203.7$ $cm$, $d_0 = 631.2$ $cm^2 year^{-1}$,
    $d_1 =4842.1$ $cm^4 year^{-1}$. These parameters are appropriate for the
    marine plant \textit{Posidonia oceanica} \cite{ruiz2017fairy,ruiz2019ABDaprox}.}
    \label{phasediagram}
\end{figure}

The location in parameter space of the different spatially
uniform states of this version of the reduced model is
summarized in Fig.~\ref{phasediagram} together with their
stability properties and is the same in both one and two
dimensions. In addition to qualitative agreement with the
dynamical regimes of the full ABD model \cite{ruiz2017fairy},
version I of the reduced model reproduces to a high degree of
accuracy the position in parameter space of the modulational
instability (MI) of the homogeneously populated solution. This
description thus provides quantitatively accurate results,
while providing a simplified model of clonal plant growth.

\subsection{Version II: effective local description}

\begin{figure}
    \centering
    \includegraphics[width=\columnwidth]{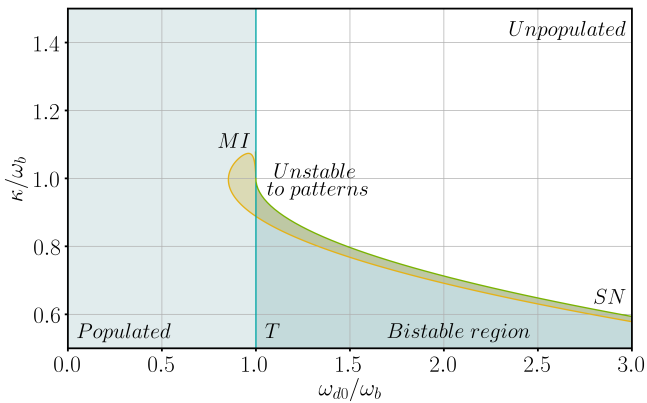}
    \caption{Phase diagram for version II of the model. The color code is as in
    Fig.~\ref{phasediagram}. The parameters are $\omega_b=0.06$ $year^{-1}$,
    $b'=12.5$ $cm^4 year^{-1}$, $a'= 100.41$ $ cm^2 year^{-1}$, $\kappa=0.048$
    $year^{-1}$, $\alpha=-8.642\cdot10^7$ $cm^6 year^{-1}$,
    $\beta=-3.585\cdot10^{13}$ $cm^8 year^{-1}$, $d_0 = 508.1$ $cm^2year^{-1}$,
    $d_1 =6560.6$ $cm^4year^{-1}$.
    }
    \label{phasediagramII}
\end{figure}

A second level of approximation, \emph{version II}, results
from performing a moment expansion of the integral term
$\omega_d[n]$ and truncating at the lowest possible order.
Specifically, we first expand the exponential inside the
integral and then truncate the moment expansion of the kernel
at fourth order. The approximation yields qualitative agreement
with the behavior of version I (and thus of the ABD model in
\cite{ruiz2017fairy}) in terms of the observed dynamical
regimes (compare Fig.~\ref{phasediagramII} with
Fig.~\ref{phasediagram}). In particular, the general ordering
of the different phases in parameter space and the nature of
the bifurcations (saddle-node, transcritical, and the MI)
bounding them is the same. But there is lack quantitative
agreement. The inclusion of higher order terms in the expansion
may improve accuracy but implies loss of simplicity. We
therefore choose the parameters in version II to preserve as
much as possible the behavior of version I while keeping its
simple form. The mortality now reads
\begin{equation}
\label{E4}
\omega_d(n)= \omega_{d0} + a'(\kappa-\omega_b) n + b' n^2 -
\alpha \partial_x^2 n - \beta \partial_x^4 n,
\end{equation}
with the intrinsic mortality $\omega_{d0}$ the same as before.
The second coefficient $a'(\kappa-\omega_b)$ controls the
degree of bistability. We write this coefficient in this way to
facilitate a comparison with version I. The coefficient $b'$
determines the saturation level, while $\alpha$ and $\beta$
come from the expansion of the nonlocal term and are
responsible for the presence or absence of spatial patterns.
The parameters $a'$ and $b'$ are chosen to generate a
bifurcation diagram similar to that of version I. The
conditions imposed are: (i) having the same density of shoots
at $\omega_{d0}=\omega_b$ and (ii) having the saddle-node
bifurcation of the homogeneous populated state at the same
value of the mortality rate $\omega_{d0}$. These conditions are
imposed at the value $\kappa=0.048 year^{-1}$ which will be
used throughout the paper. The parameters $\alpha$ and $\beta$
are chosen to generate the modulational or Turing instability at a
similar mortality rate as in version I for the same chosen
value of $\kappa$, and with a similar critical wavelength.

The equation for this version II of the model, resulting from
combining Eqs. (\ref{E1}) and (\ref{E4}), was originally
derived in \cite{ruiz2019ABDaprox}, and some of its properties
studied in two dimensions. It has several terms in common with
previously studied models. In particular the effect of the
nonvariational terms $\left|\partial_x n\right|^2$ and
$n\partial_x^2 n$ on front motion was considered in
\cite{AlvarezSocorro2017nonvariational} in a model of an
optical system. In the context of vegetation patterns, the
model discussed in
\cite{FernandezOto2019front,ParraRivas2020formation} is very
similar to our version II, although it misses the
$\left|\partial_x n\right|^2$ term, which is characteristic of
clonal growth.

\section{Stationary patterns and localized structures}

In this section we discuss the different stationary solutions
supported by versions I and II of the model in one dimension.
We first show the results for version I with the full nonlocal
interaction term, and then for version II based on the
truncated moment expansion, highlighting the main differences
between them. Throughout we use the mortality rate
$\omega_{d0}$ as the main control parameter. In order to follow
stationary solutions we use a pseudo-arclength continuation
method \cite{Keller1988,Mittelmann1986} where the Jacobian is
calculated in Fourier space. Starting with an initial condition
obtained using numerical simulations we continue the stable and
unstable branches changing $\omega_{d0}$ as a control parameter.

\subsection{Version I: patterns and localized structures}

\begin{figure}
  \includegraphics[scale=0.4]{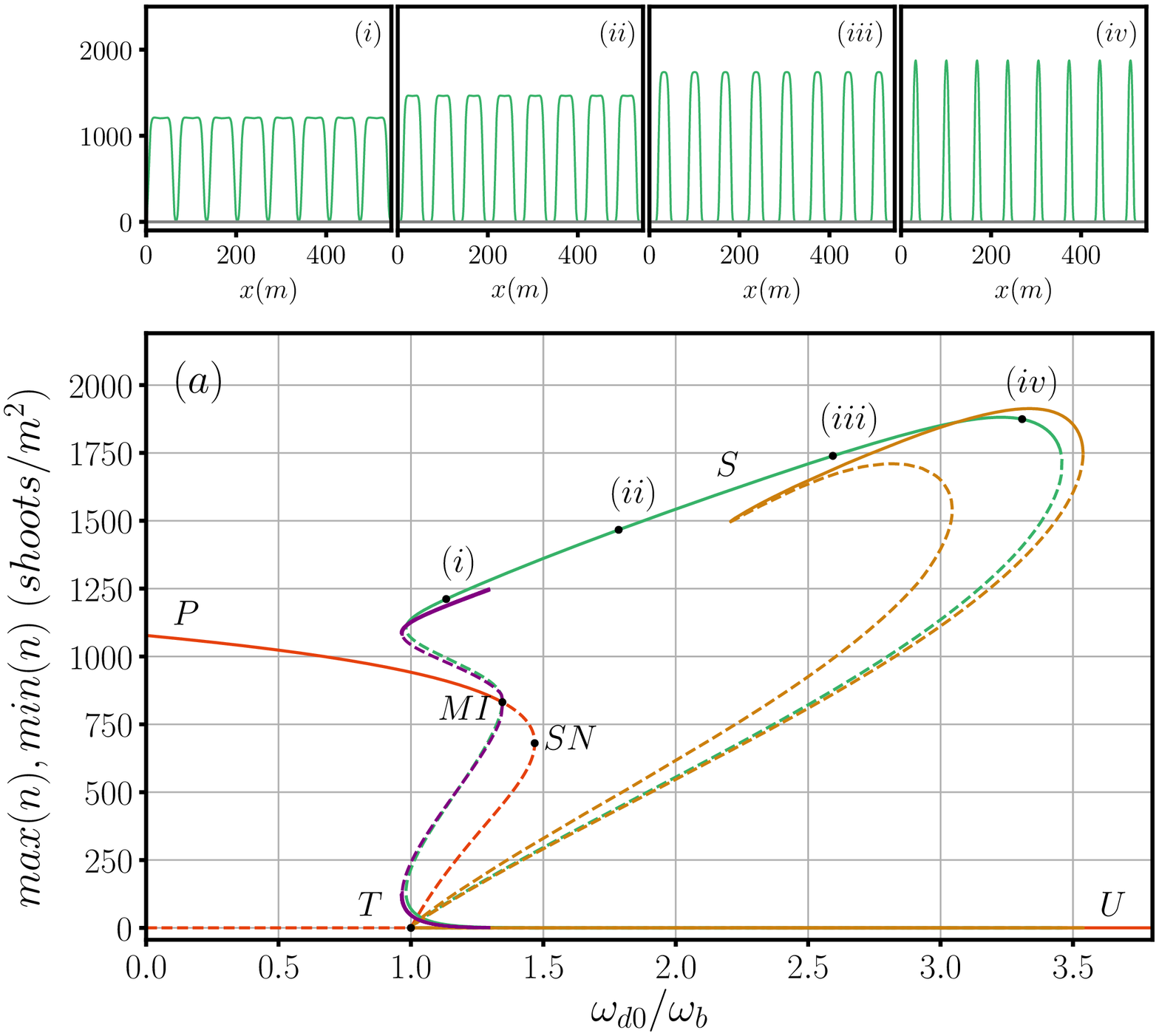}
  \includegraphics[scale=0.378]{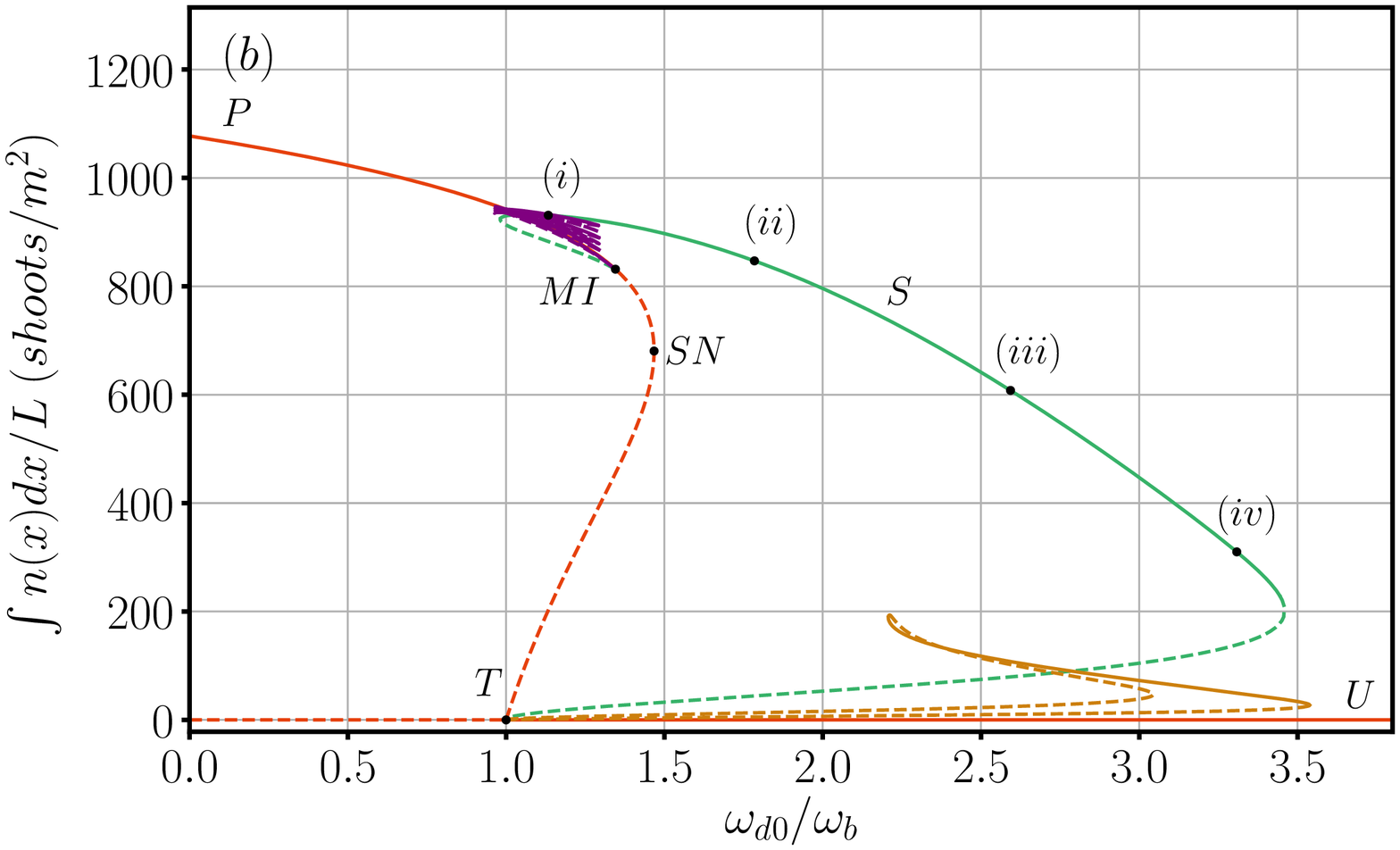}
\caption{Bifurcation diagram for version I of the model as a function of
the ratio $\omega_{d0}/\omega_b$. Panel (a) shows the maximum
and minimum values of the density $n$ corresponding to the
different stationary states while (b) shows the average density.
Continuous (dashed) lines represent stable (unstable) solutions.
The P and U states are shown in red, while the spatially periodic
stripe state S is shown in green (sample S states are shown in the
panels above (a)). Spatially localized states (LS) corresponding
to holes embedded in the P state are shown in purple while the orange
curves show localized structures consisting of patches of vegetation
on bare soil, i.e., embedded in the U state. Here $\kappa=0.048$ $year^{-1}$
and the remaining parameters are as in Fig.~\ref{phasediagram}.
}
\label{Bifdiagker}
\end{figure}

Different solutions are observed when the mortality
$\omega_{d0}$ changes as summarized in the bifurcation diagram
shown in Fig.~\ref{Bifdiagker}. When mortality is large the
branching rate $\omega_b$ is insufficient to sustain growth and
the only stable solution is bare soil, the unpopulated state
(U). In contrast, when the death rate is small compared to the
branching rate the homogeneous populated state (P) prevails. In
between one finds a region of coexistence between P and U; this
region terminates in a saddle-node bifurcation labeled $SN$.
Both the populated and unpopulated states are shown in red in
the figure. When the branching and mortality rates are
comparable the upper P state may become unstable to spatial
modulations that develop into a periodic pattern that we call a
\textit{stripe pattern} (S), shown in green in
Fig.~\ref{Bifdiagker}. The emerging stripe pattern bifurcates
subcritically but undergoes a fold, thereby generating a region
of coexistence between stable stripes and the stable upper P
state for mortalities below MI. The stripe pattern turns out to
be rather robust and stable stripes are found far beyond the
region of existence of the homogeneous state, coexisting with
the bare soil state U over a broad range of values of
$\omega_{d0}$ above the transcritical bifurcation $T$ of U.
With increasing mortality rate the stable stripes eventually
terminate in a fold bifurcation. The unstable S states that
result in turn terminate at a second MI or Turing bifurcation
located on the unstable (middle) branch of P, very close to
zero density (Fig.~\ref{Bifdiagker}b). We mention that between
these two Turing bifurcations there are other pairs of Turing
bifurcations that also give rise to spatially periodic stripes
but with wavelengths different (smaller and larger) from the
critical wavelength corresponding to the MI of the P state,
which is the one displayed here. Thus, the S state is by no
means unique.

\begin{figure}
  \includegraphics[width=1\columnwidth]{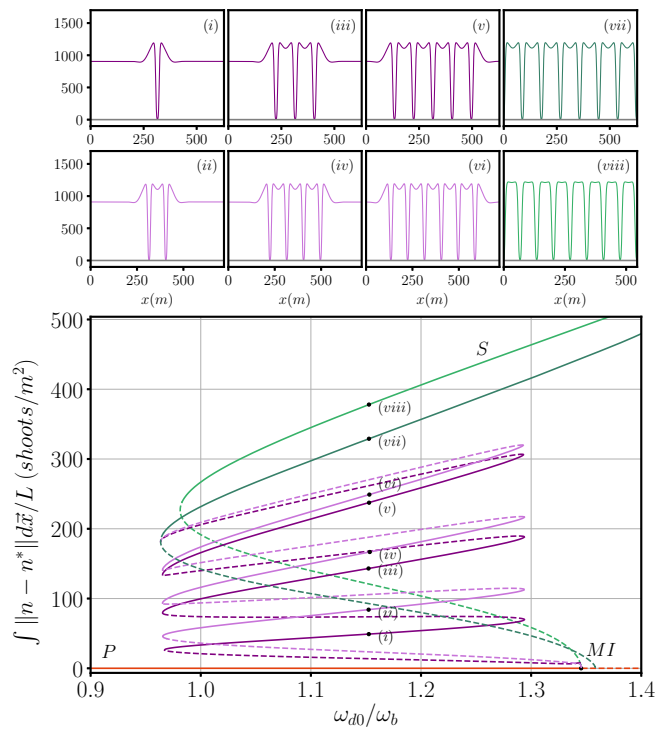}
\caption{Close-up of the snaking region of version I of the model, emerging from the modulational
instability of the P state in Fig.~\ref{Bifdiagker} using the norm of
the difference between this state and P to reveal details of the snaking
bifurcation diagram. The dark purple curve represents LS with an odd number
of holes, while the bright purple curve represents LS with an even number.
The green curves correspond to stripe patterns S with two different
wavenumbers, one with the critical wavenumber and the other with the wavenumber
selected by a stationary front between the homogeneous and pattern states,
which determines the wavelength within the LS. The upper panels show the
solution profiles corresponding to the labeled locations in the bifurcation
diagram (lower panel). The parameters are as in Fig.~\ref{Bifdiagker}.}
\label{snaking}
\end{figure}

Figure~\ref{Bifdiagker} shows that in addition to the S branch,
the MI or Turing bifurcation on the upper P branch generates a
pair of branches of spatially localized structures (LS) that
also emerge subcritically, creating a window in the mortality
rate, called a snaking region \cite{Woods99,Coullet00,Burke06},
within which one finds stable stationary states consisting of
segments of the periodic S state of arbitrarily large length,
embedded in the background P state. The purple lines in
Figs.~\ref{Bifdiagker} and \ref{snaking} show the resulting
snaking bifurcation diagram revealing the presence of two
intertwined LS branches consisting of states with odd and even
numbers of close-packed troughs. A single-trough state corresponds
to a single region of nearly bare soil embedded in P, i.e., a hole in
an otherwise homogeneous state, analogous to fairy circles in
two spatial dimensions. Based on the general theory developed
for the prototypical Swift-Hohenberg equation we expect that
opposing folds on the odd and even branches are connected by
(unstable) branches of asymmetric states. Owing to the nonvariational
structure of the present problem we expect that these
states drift, cf.~\cite{Mercader2013}. In this work we do not
follow unstable states of this type. We also note that the
region of existence of LS extends beyond the left fold of the
S branch shown in the figure. This is possible when the wavenumber
selected by the LS in the snaking region differs sufficiently from
the critical wavenumber at MI to force the LS branches to terminate
on a different S branch (see Fig.~\ref{snaking}). In the present
case the resulting snaking region extends almost to the fold of
this second S branch indicating that the chosen parameter
values are very close to a transition that breaks up the snaking
scenario. This transition occurs when the left LS folds touch the
left fold of the corresponding pattern state~\cite{Wagenknecht12}.

\begin{figure}
  \includegraphics[width=1\columnwidth]{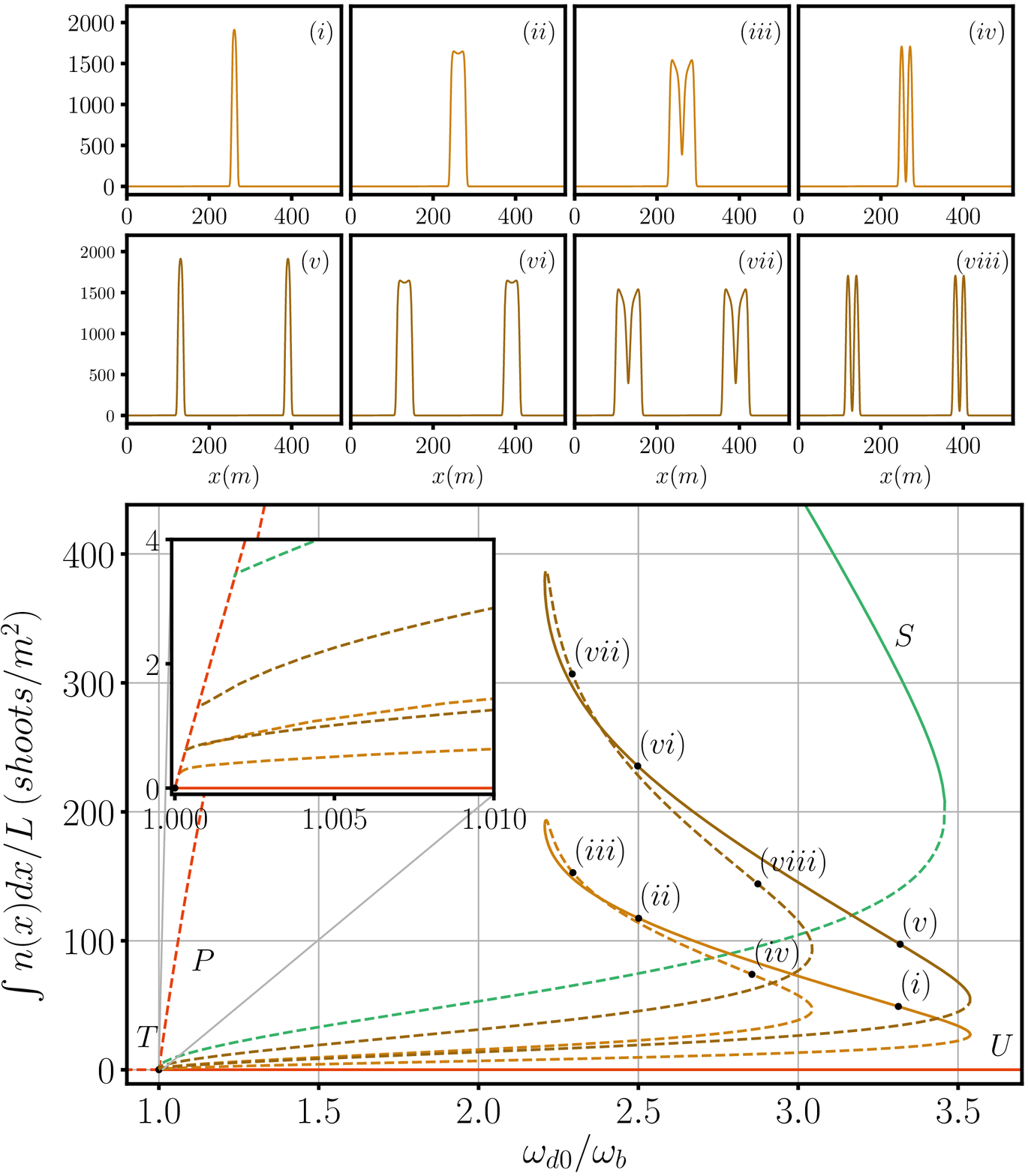}
  \caption{Close-up of the region of vegetation spots on bare soil from version I of the model.
  Light orange curve represents LS with one peak, while dark orange
  represents LS with two peaks separated by half the system size.
  The stripe pattern S with the critical wavenumber is shown in green.
  The upper panel shows the solution profiles corresponding to the
  labeled locations in the bifurcation diagram (lower panel). The
  parameters are as in Fig.~\ref{Bifdiagker}.}
\label{snaking_spot}
\end{figure}

For mortalities larger than the branching rate,
$\omega_{d0}>\omega_b$, a different type of stable spatially
localized structure is found, consisting of isolated vegetation
patches on bare soil (see orange lines in
Figs.~\ref{Bifdiagker} and \ref{snaking_spot}). The bifurcation
structure of these LS differs from the previous case, and they
do not lie on a standard snaking branch. Instead these states
bifurcate from the unstable P branch close to zero density and
below the MI bifurcation at which the S state terminates (see
inset in Fig.~\ref{snaking_spot}). This bifurcation corresponds
to a long wavelength instability, and generates a state with
wavelength equal to the system size. The low-amplitude
one-patch solution that arises is unstable and grows in
amplitude with increasing mortality until it reaches a fold
where it acquires stability and becomes a high amplitude stable
single patch solution. Beyond the fold the patch state
continues to grow in amplitude but now for decreasing mortality
until it reaches a second fold. Near this second fold the
solution starts to change shape, the central part of the patch
developing a relative minimum that continues to decrease along
the subsequent branch. Thus, the patch starts to divide into
two patches until, after a third fold, two low-amplitude LS are
present and these gradually decrease in amplitude with
decreasing mortality until the branch terminates back on the
unstable P branch. Very close to this bifurcation both peaks
become very shallow and their maxima move rapidly apart until
they are separated by $L/2$, i.e., half the system size. Thus,
the termination point corresponds to a pattern-forming
bifurcation of the unstable P state to a two-peak state much
like the long-wave bifurcation to the single peak state that
occurs at a lower value of the density. The bifurcation to the
two-peak state is in fact a pitchfork bifurcation, one fork of
which corresponds to the termination of the two-peak state
generated from the single peak state via peak-splitting as just
described, while the other takes part in a similar scenario but
based on a two-peak state. This scenario results, again via
peak-splitting, in a four-peak state that terminates on P at
yet higher (but still small) density (Fig.~\ref{snaking_spot}).
Once again, close to this termination point the four peaks
become equidistant, i.e., separated by $L/4$, allowing this
branch to terminate in a pattern-forming bifurcation from a
periodic state. In fact this behavior is observed for any
number of equispaced identical peaks, even or odd, generated in
corresponding bifurcations along the unstable P branch. Similar
bifurcation structures have been found in other systems
\cite{Yochelis08,Parra2018}. We conjecture that in the limit of
an infinite domain the wavelength along the unstable S branch
increases by wavelength doubling that occurs via the same
process as that occurring for the one-peak and two-peak states,
i.e., via repeated peak-splitting \cite{Yochelis08,Parra2018},
ultimately reaching the transcritical bifurcation T and zero
wavenumber, much as occurs in the Gray-Scott model
\cite{Doelman1998,Doelman2012}. This scenario is supported by
the fact that in all cases the folds on the right align at
$\omega_{d0}/\omega_b\approx 3.538$, a value that is close to
that of the fold on the S branch, while the folds on the left
align at $\omega_{d0}/\omega_b\approx 2.207$; the intermediate
folds are also aligned (at $\omega_{d0}/\omega_b\approx
3.044$). Ref.~\cite{Zelnik18} describes a scenario whereby the
single peak state may reconnect with or turn into a pattern
state.

The stationary solutions shown have been computed using
different discretizations $\Delta x$. Periodic solutions have
been computed using a domain size equal to one wavelength and
have been represented showing multiple wavelengths to emphasize
the extended nature of the pattern. For the stripes with
critical wavenumber shown in Figs.~\ref{Bifdiagker},
\ref{snaking} and \ref{snaking_spot}, the number of grid points
is $N = 128$ and $\Delta x = 0.533$ $m$. For stripes with
wavelength selected by the localized structure shown in
Figs.~\ref{snaking} $N = 128$ and $\Delta x = 0.703$ $m$. For
all LS $N=1024$. Those with odd and even number of holes shown
in Figs.~\ref{Bifdiagker} and \ref{snaking} use $\Delta x =
0.615$ $m$ and $\Delta x = 0.703$ $m$, respectively. For
patches shown in Figs.~\ref{Bifdiagker} and \ref{snaking_spot},
$\Delta x = 0.509$ $m$.

\subsection{Version II: patterns and localized structures}

\begin{figure}
  \includegraphics[scale=0.38]{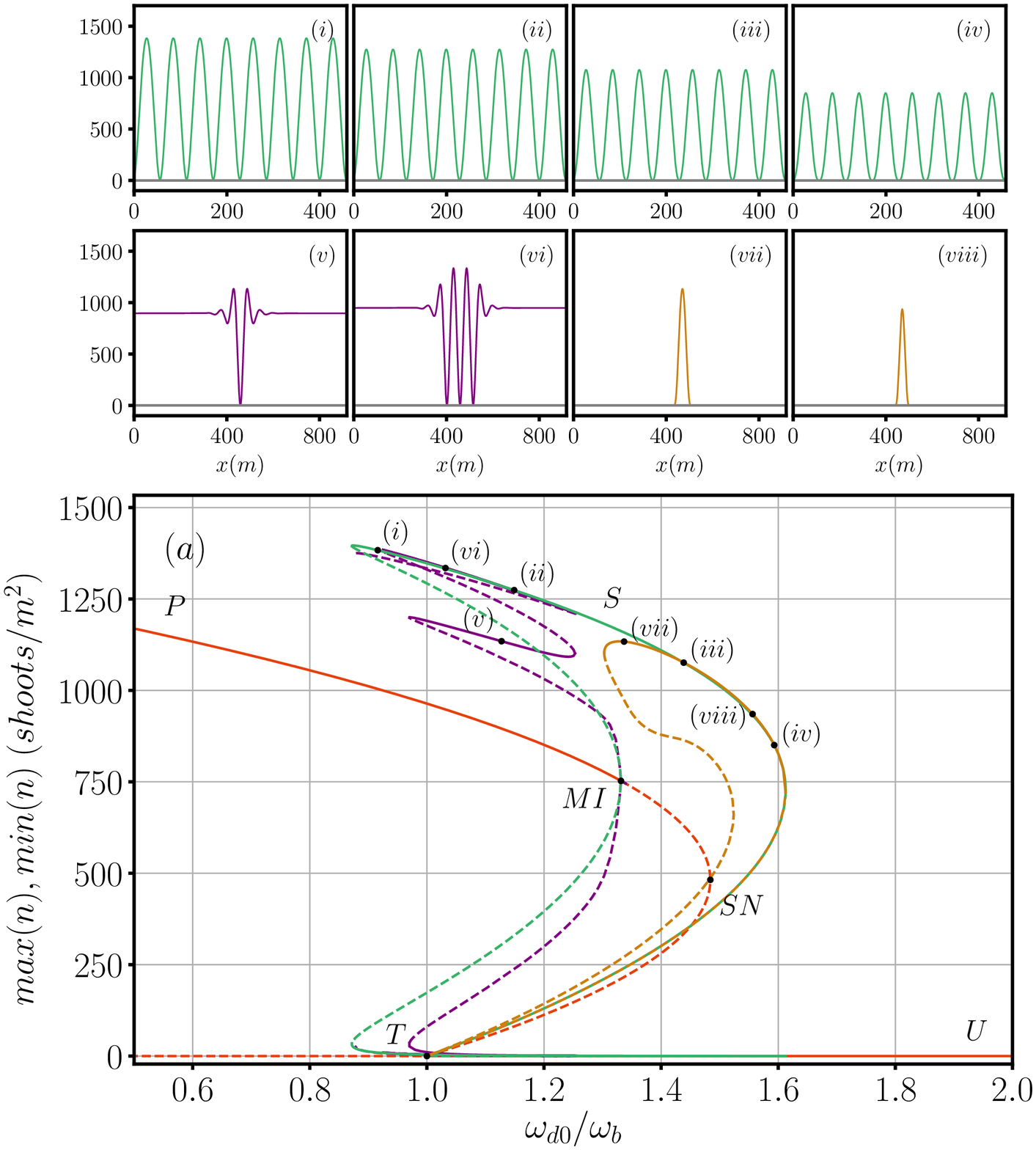}
  \includegraphics[scale=0.38]{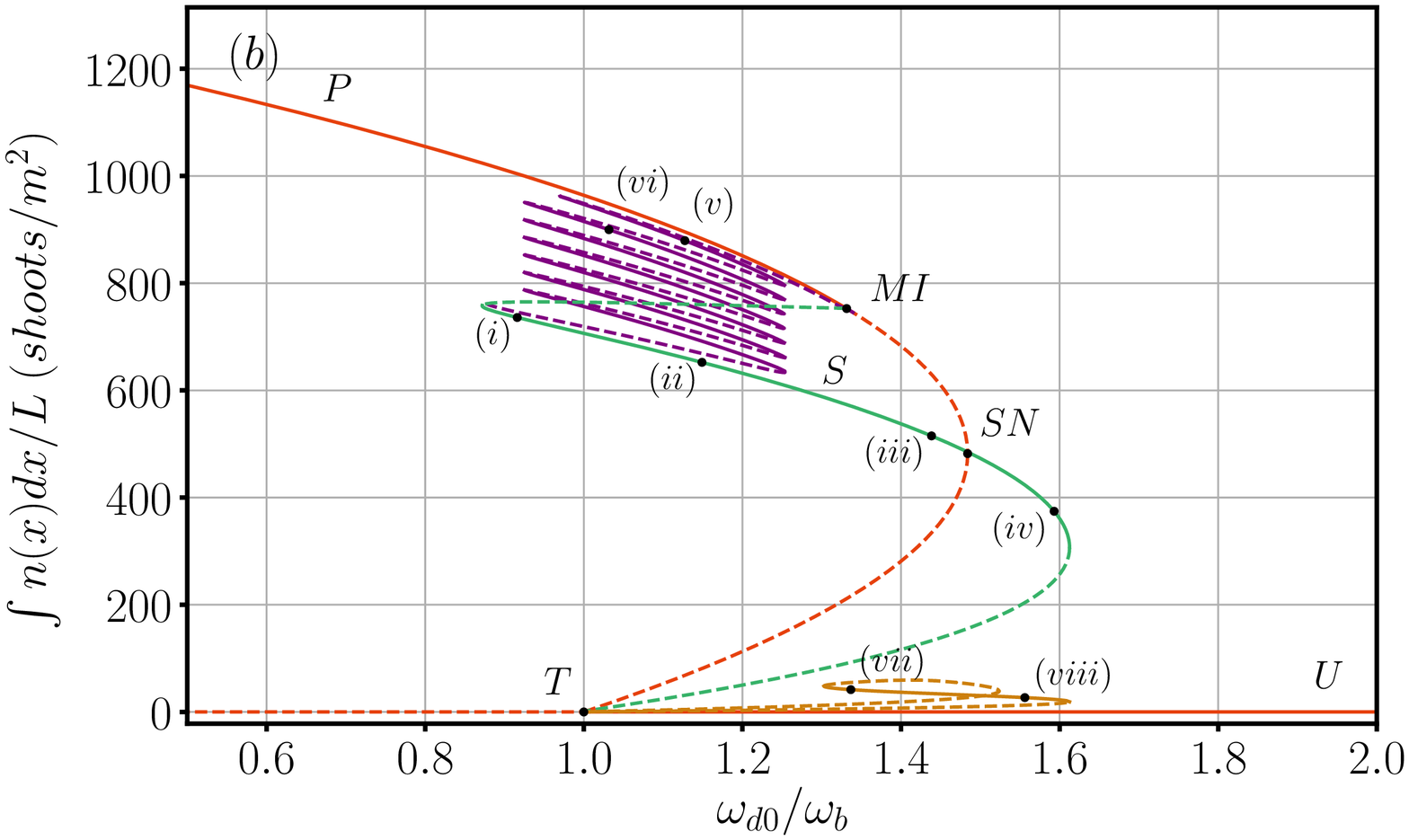}
\caption{Bifurcation diagram and sample stationary solutions
of version II as a function of the ratio $\omega_{d0}/\omega_b$.
The color scheme, labels and line types are as in Fig.~\ref{Bifdiagker};
$\kappa=0.048$ $year^{-1}$ and the remaining parameters are as in
Fig.~\ref{phasediagramII}.}
\label{Bifdiagsimp}
\end{figure}

Figure \ref{Bifdiagsimp} shows the corresponding bifurcation
diagram for the P, U, S and LS states in version II of the
model. The bifurcation scenario is qualitatively similar to
that observed in version I (compare Fig.~\ref{Bifdiagsimp} with
Fig.~\ref{Bifdiagker}), confirming the fact that this simpler
version of the model captures all the basic mechanisms.
However, substantial quantitative differences are observed. For
instance, the mortality ranges in which each solution exists
are reduced, while the solution profile becomes more
triangular. As a result the bare soil minima in the S state are
much narrower.

These solutions have been computed following the same procedure
as for version I of the model. For the stripes $N =128$, while
for all localized structures $N =2048$, with $\Delta x = 0.448$
$m$ in all cases.

\section{Fronts}

Figure \ref{Bifdiagker} for version I and Fig.
\ref{Bifdiagsimp} for version II reveal the existence of
several different regions of coexistence between the U, P and S
states owing to the presence of multiple stable solutions,
raising the possibility of a number of different fronts
connecting these states. As many as three stable spatially
extended states can coexist simultaneously, a situation that
also arises in other vegetation pattern-forming models
\cite{Zelnik18}. Here we study the fronts connecting the
populated state with the unpopulated state (P-U fronts), the
stripe pattern with the populated state (S-P fronts) and the
stripe pattern with the unpopulated state (S-U fronts). We note
that our one-dimensional study is also relevant for flat fronts
in two dimensions advancing in the normal direction.

However, flat fronts may experience transversal instabilities
in two dimensions changing the front profile (see e.g.
\cite{FernandezOto2019front}). From this perspective, a
dedicated investigation is needed to achieve a complete
description of front propagation in two dimensions.
Nevertheless, the propagation of fronts is strongly influenced
by the stability range of LS, where the one- and
two-dimensional cases show general agreement
\cite{ruiz2019ABDaprox}.

Front dynamics depend strongly on the stability of the states
that are involved. The speed of moving fronts connecting two
linearly stable states necessarily depends on nonlinear
processes, i.e. processes occurring beyond the immediate
vicinity of the front. Such fronts are called \emph{pushed}
\cite{VanSaarloos2003}. In contrast, fronts of constant form
describing the invasion of a linearly unstable state by a
stable one may travel with a speed determined via a linear
mechanism that requires that perturbations ahead of the front
grow at just such a rate that a front of constant form is
maintained~\cite{VanSaarloos2003}. These fronts are thus
\emph{pulled} by the linear instability ahead of them. It
should be mentioned, however, that the existence of pulled
fronts does not imply that such fronts are selected. Pushed
fronts can exist in the same parameter regime and these would
travel with a different speed. In the case of coexistence
between pulled and pushed fronts, the front with the larger
velocity is usually the one that is observed
\cite{Hari,Archer}.

The speed $v$ of a pulled front can be obtained by considering
infinitesimal perturbations, of the form $e^{ikx+\lambda
\left(k\right) t}$ or equivalently $e^{ikx'+\Lambda (k) t}$
with $\Lambda(k)=ikv+\lambda(k)$, to the unstable state in the
comoving reference frame $x'=x-vt$
\cite{,VanSaarloos2003,VanSaarloos1993}. Applying the condition
of \emph{marginal stability}, i.e., that in the frame moving at
speed $v$ perturbations neither grow nor decay, leads to the
requirement that the (complex) group speed and the growth rate
of the perturbation both vanish, yielding three conditions that
are solved for the unknown speed $v$ of the front and the real
and imaginary wavenumbers at the leading edge of the front.
Specifically, the conditions are $\textrm{Re}\left[\Lambda
(k)\right]$=0 and $d\Lambda (k)/dk = 0$, where $k=k_r+ik_i$,
the real and imaginary parts of $k$ representing the wavenumber
ahead of the front and the spatial decay of its envelope. These
equations can be written as $v=\textrm{Re}\left[\lambda
(k)\right]/k_i$, $\textrm{Re}\left[d\lambda (k)/dk\right]=0$
and $v = -\textrm{Im}\left[d\lambda (k)/dk\right]$. This
calculation is further developed in the Appendix, and applied
to specific fronts in our model systems as described in the
following sections.

\subsection{Fronts in version I of the model}
\label{results_modelI}

In this section we show the results obtained with version I,
using the bare mortality $\omega_{d0}$ as the main control
parameter and keeping the other parameters as in
Fig.~\ref{Bifdiagker}. Our numerical simulations use a
pseudospectral method with periodic boundary conditions and
$\Delta t=1.667 \cdot 10^{-3}$ $years$, $\Delta x=0.255$ $m$
and $N=4096$, $8192$ and $16384$ grid points starting with a
homogeneous initial condition, the U or P state, with a very
narrow step function at $x=0$ to excite a competing solution.
In the cases in which two distinct fronts are possible
(tristability) we use the profile of the desired front obtained
for other values of the mortality as initial condition.

\begin{figure}
  \includegraphics[width=1\columnwidth]{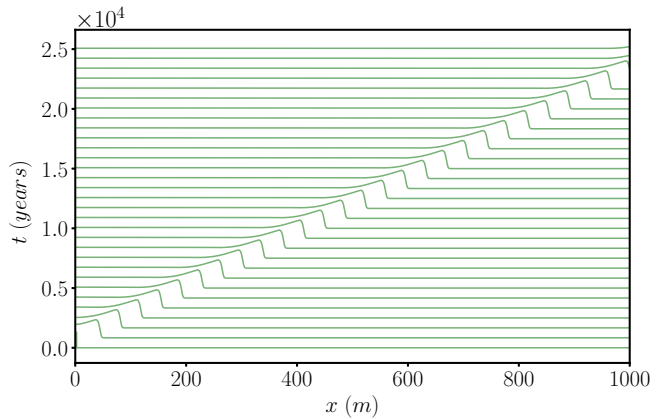}
  \caption{Space-time representation of a pushed P-U front whereby a stable
P state invades a stable U state in version I of the model. The figure
shows $n(x,t)$ along the vertical axis at successive times $t$, displaced in the vertical
by $t$. The front travels with speed $v\approx$ $0.044$ $m/year$. Here
$\omega_{d0}=0.0679$ $year^{-1}$ ($\omega_{d0}/\omega_b=1.132$) and the remaining parameters
are as in Fig.~\ref{Bifdiagker}.}
\label{frontP-U}
\end{figure}

We study first the P-U fronts between the two homogeneous
solutions. We can distinguish two cases. When
$\omega_{d0}/\omega_b>1$ (but below a value at which the state
P behind the front destabilizes, see below) the front is a
pushed front as both P and U are stable. The front speed as
well as the direction of advance is thus determined by
nonlinearities. We observe numerically that P always invades U.
Figure~\ref{frontP-U} shows an example of this type of front.
On general grounds, one would expect to find a sufficiently
large value of the mortality, a Maxwell point, beyond which the
direction of the front reverses and the bare soil invades the
populated solution. However, it turns out that this occurs
beyond the mortality rate for the instability of the P solution
to pattern formation (MI), and we never observe this type of
desertification front.

\begin{figure}
    \centering
    \includegraphics[width=1\columnwidth]{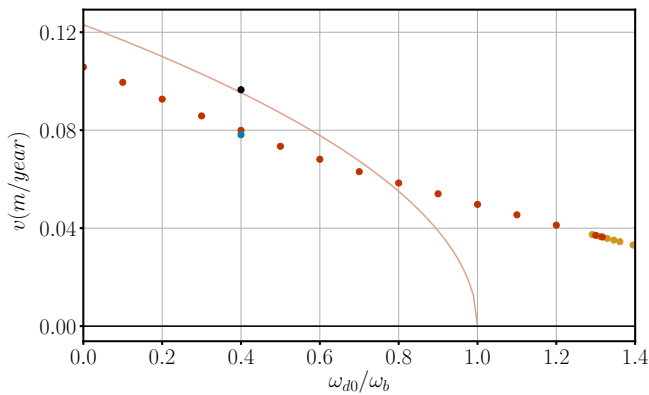}
    \caption{Speed $v$ of the P-U (red dots) and S-U (yellow dots)
      fronts in version I of the model as a function of $\omega_{d0}/\omega_{b}$.
      Solid line represents the linear marginal stability prediction for a pulled P-U
      front. The blue dot corresponds to a numerical simulation without the term
      $\left|\partial_x n\right|^2$, while the black dot corresponds to removing
      in addition all nonlocal competition ($\sigma_\kappa=0$). Parameters are as
      in Fig.~\ref{Bifdiagker}.
}
    \label{velocityP-U}
\end{figure}

 For $\omega_{d0}/\omega_b<1$ the U state is unstable
and a pulled front whereby P advances into U exists. Its speed
can be computed from the marginal stability approach. In
Fig.~\ref{velocityP-U} we show, as a function of
$\omega_{d0}/\omega_b$, the speed $v$ of the pulled front
computed analytically (red line, see Appendix) and from
numerical simulations (red dots). It is clear that the marginal
stability prediction of the speed fails. The front speed for
$\omega_{d0}/\omega_b<1$ appears to be a continuation of the
pushed front speed for $\omega_{d0}/\omega_b>1$. We have
investigated the discrepancy between the linear marginal
stability prediction and numerics by performing numerical
simulations in which the nonlinear terms that do not appear in
the linear calculation are removed. First we removed the term
$d_1 \left|\partial_x n\right|^2$ but the resulting change in
the speed of the front is small (blue dot in
Fig.~\ref{velocityP-U} for $\omega_{d0}/\omega_b = 0.4$). We
then removed the nonlocal competition term as well by setting
$\sigma_\kappa=0$ (black dot in Fig.~\ref{velocityP-U}). In
this second case the velocity changes dramatically and
coincides with the linear marginal stability prediction. This
result points to the nonlocal interaction term as the source of
discrepancy between the linear marginal stability prediction
and the speed of the front in the full system. We ascribe this
effect to an inhibiting effect of existing plants at the edge
of the front on the growth of plants a certain distance ahead
of the front, thereby decreasing the speed of propagation. This
interpretation is supported by the presence of a density
maximum in the front profile located at the front edge and
generated by long range interactions (Fig.~\ref{frontP-U}). We
note that in systems of reaction-diffusion equations the speed
selection problem is more complex (the fastest front is not
always the one selected) than in single species problems and
nonlocal systems such as that studied here are, somewhat
loosely, equivalent to a higher-dimensional
system~\cite{Scheel}. Other effects of nonlocal terms on the
speed of fronts have been studied
in~\cite{Gelens10,FernandezOto13,FernandezOto14}.

The P-U front is observed for $\omega_{d0}/\omega_b$ below a
critical value $1.325\pm 0.008$  (which is close to but below the
MI occurring at $\omega_{d0}/\omega_b\approx 1.345$). In this
region P always invades U and no Maxwell point (i.e., a value of
$\omega_{d0}$ at which propagation direction reverses) is
found. When $\omega_{d0}/\omega_b\gtrsim 1.325\pm0.008$, the P
state behind the front destabilizes so that the front generates
a stripe pattern in its wake, i.e., it becomes a S-U front. We
note, however, that there is a small region
($1.277\pm0.014\lesssim\omega_{d0}/\omega_b\lesssim 1.325\pm0.008$)
of coexistence between the two fronts (yellow dots in Fig.~\ref{velocityP-U}).

\begin{figure}
  \includegraphics[width=1\columnwidth]{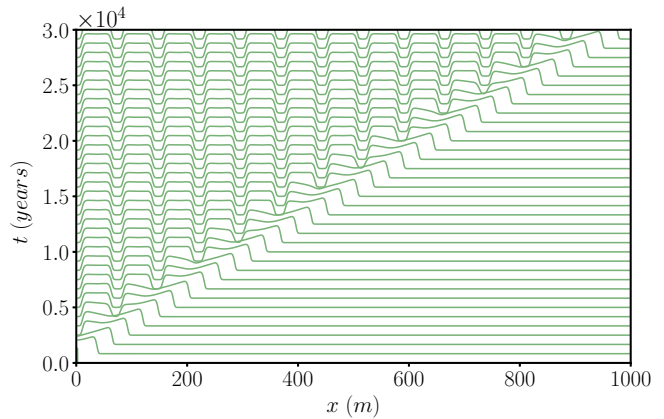}
  \caption{Space-time representation of a pushed S-U front connecting a
stable stripe pattern S with the unpopulated state U in version I of the model.
The front travels with speed $v\approx 0.033$ $ m/year$; $\omega_{d0}=0.0837$ $year^{-1}$
($\omega_{d0}/\omega_b=1.395$) and the remaining parameters are as in Fig.~\ref{Bifdiagker}.}
\label{frontS-U}
\end{figure}

\begin{figure}
  \includegraphics[width=1\columnwidth]{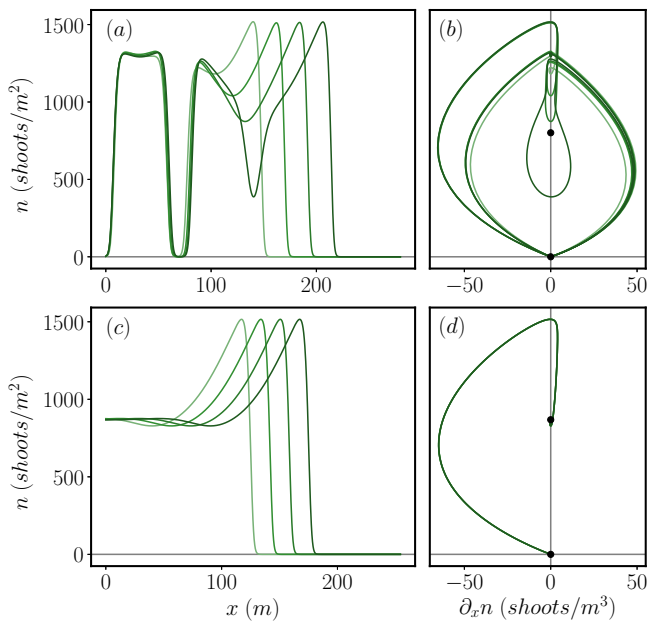}
  \caption{Spatial profiles of S-U and P-U fronts at equispaced times.
  Panel (a) shows the spatial profile of the density $n(x)$ across an S-U
  front and (b) its appearance in a $(\partial_xn,n)$ plot, both for
  $\omega_{d0}=0.0837$ $year^{-1}$ ($\omega_{d0}/\omega_b=1.395$). The black dots in panels (b)
  indicate the bare soil
  state U and the homogeneous populated state P. Panels (c) and (d) show the corresponding
  quantities for a P-U front at $\omega_{d0}=0.0758$ $year^{-1}$ ($\omega_{d0}/\omega_b=1.263$). The remaining
  parameters are as in Fig.~\ref{Bifdiagker}.}
\label{heteroclinics}
\end{figure}

\begin{figure}
  \includegraphics[width=1\columnwidth]{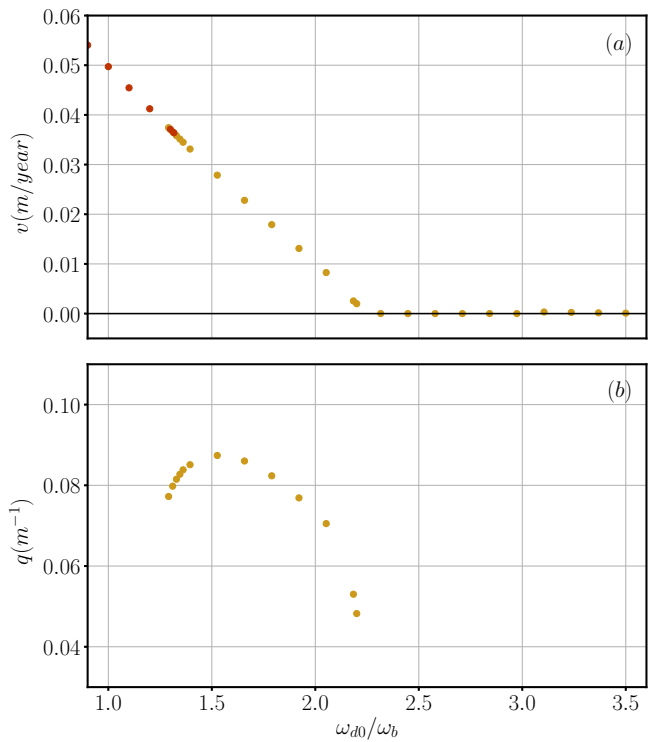}
\caption{Speed $v$ and wavenumber $q$ of the S-U (yellow dots) front and speed $v$ of the P-U (red dots)
front as a function of $\omega_{d0}/\omega_{b}$ in version I of the model.
In panel (a) the speeds of the two fronts fall on the same curve, although hysteresis is present
in the interval $1.277\pm0.014\lesssim\omega_{d0}/\omega_b\lesssim1.325\pm0.008$. In panel (b) a nonzero wavenumber is selected in the interval $1.291\lesssim \omega_{d0}/\omega_{b}\lesssim 2.207$, i.e., in the
interval within which the S-U front travels and is stable.
Parameters as in Fig. \ref{Bifdiagker}.}
\label{velocityS-U}
\end{figure}

Figure \ref{frontS-U} shows a space-time representation of an
S-U front at $\omega_{d0}/\omega_b =1.395$. This front, whereby
S invades U, is pushed since both S and U are stable at this
mortality rate. The front's leading edge is very steep, and is
followed by a sloping plateau that leads to the formation of a
deep hole that is characteristic of the S state at this
mortality rate. Once the hole forms the plateau relaxes,
reproducing the S profile near its maxima. Note that in the
reference frame moving with the front the deposition of
successive holes is an oscillatory process with a well-defined
{\it temporal} period and we conjecture that, in that reference
frame, the S-U front forms via a (subcritical) Hopf bifurcation
of the P-U front. For this reason the leading edge of the S-U
front at, say, $\omega_{d0}/\omega_b \sim 1.395$ closely
resembles that of the corresponding P-U front
(Fig.~\ref{heteroclinics}), a fact that is likely responsible
for the similar speeds of these two fronts
(Figs.~\ref{velocityP-U} and \ref{velocityS-U}). Indeed,
Figs.~\ref{heteroclinics}(a,c) show the profiles of the S-U and
P-U fronts at equispaced times, with (a) showing the initiation
of the deep hole associated with the S state. No such hole is
generated behind a P-U front. Panels (b,d) show that these
fronts can be viewed, respectively, as heteroclinic connections
between a limit cycle (the S state) and the trivial or zero
state (the U state), and between two equilibria, one
corresponding to the P state and the other to the U state,
indicated by the black spots in the figure. Note the similarity
of the trajectory leaving the zero state. In both cases the
density profile has a pronounced maximum just behind the
leading edge of the front, which we ascribe to the absence of
competition ahead of the front. Such overshoots are
characteristic of fairy circles in arid ecosystems as well.

The S-U front is observed for $\omega_{d0}/\omega_b \gtrsim
1.277\pm0.014$ and travels with a speed that decreases
monotonically with $\omega_{d0}$ until
$\omega_{d0}/\omega_b\approx 2.205\pm0.005$ where its speed
vanishes (Fig.~\ref{velocityS-U}). In this interval of
mortalities the S-U front selects a well-defined and nonzero
wavenumber $q$ in its wake, with values shown in
Fig.~\ref{velocityS-U}. Beyond $\omega_{d0}/\omega_b\approx
2.205$ no new holes are generated and a stationary state
consisting of equispaced, widely separated stripes is observed.
The stopping of the front is not the result of conventional
front pinning \cite{Coullet00}, since the spatial eigenvalues
of U cannot be complex owing to the requirement that the
density $n(x)$ is everywhere non-negative. We have been unable
to determine whether the selected wavenumber $q$ remains
nonzero at $\omega_{d0}/\omega_b\approx 2.205$ but we mention
that the region $\omega_{d0}/\omega_b\gtrsim 2.205$ is in any
case populated by a number of stable stationary
equispaced-patch states resembling the 1-peak, 2-peak localized
structures described in Sec.~III.1; only one of these would be
selected by the moving front as $\omega_{d0}/\omega_b\to
2.205$, most likely corresponding to $q=2\pi/L$ or the 1-peak
state. Thus, the transition at $\omega_{d0}/\omega_b\approx
2.205$ would correspond to a transition from a state with an
intrinsic wavelength $2\pi/q$ to one where the wavelength is
determined by the domain size $L$.

\begin{figure}
  \includegraphics[width=1\columnwidth]{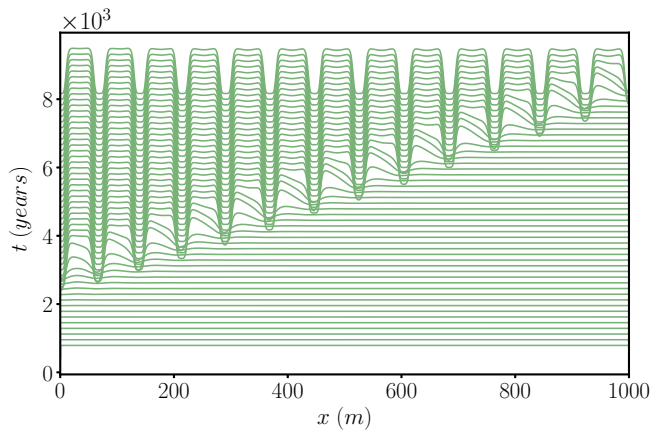}
\caption{Space-time representation of a pulled S-P front in version I of the model.
  Here stable stripes S invade an unstable homogeneous state P with speed
  $v\approx0.169$ $m/year$, which is well reproduced by the marginal stability
  calculation corresponding to a pulled front. The transient responsible for the
  initial thinner stripes decays rapidly leaving a well-defined stripe wavelength.
  Here $\omega_{d0}=0.0846$ $year^{-1}$ ($\omega_{d0}/\omega_b=1.41$) and the remaining
  parameters are as in Fig.~\ref{Bifdiagker}. }
\label{frontS-P}
\end{figure}

\begin{figure}
  \includegraphics[width=1\columnwidth]{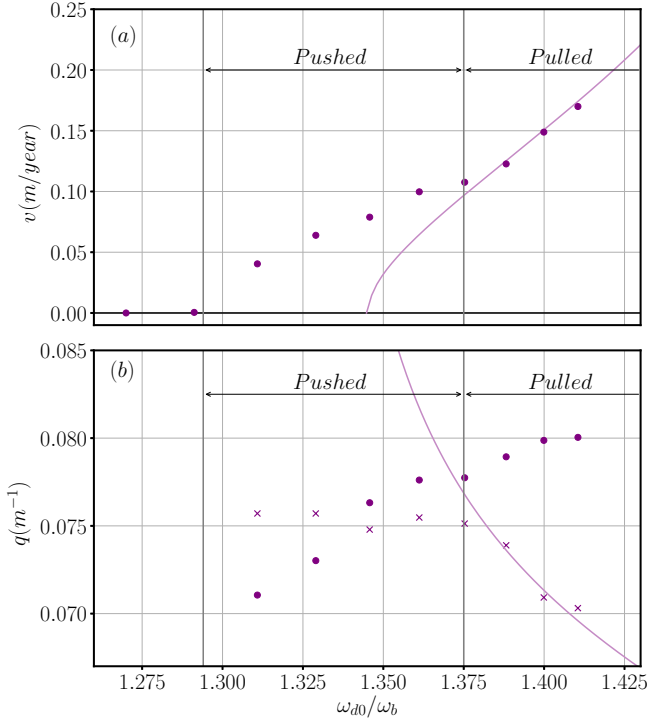}
  \caption{Speed $v$ and wavenumber $q$ of an S-P front in version I of the model
    as a function of $\omega_{d0}/\omega_{b}$. Dots and crosses correspond to
    numerical simulations while the solid line represents the linear marginal
    stability prediction for a pulled front. In panel (b), dots correspond to the
    wavenumber of the S state deposited behind the front while crosses correspond to the
    wavenumber measured at the leading edge. The prominent change in slope, identified by a
    vertical line, is associated with the transition from a pushed front to a pulled front.
    Parameters are as in Fig.~\ref{Bifdiagker}.}
\label{velocityS-P}
\end{figure}

The last front we consider is an S-P front between the stripe
pattern S and the populated state P. The stability of the P
state changes at the MI instability (occurring at mortality
$\omega_{d0,c}$, with $\omega_{d0,c}/\omega_b = 1.345$),
whereas S is always stable in the region of coexistence with P.
For $\omega_{d0}<\omega_{d0,c}$) the only possible front is
pushed since both S and P are stable. As
$\omega_{d0,c}/\omega_b$ decreases from 1.345 the speed of the
pushed front also decreases and falls to zero at the right
boundary of a pinning region which extends from
$\omega_{d0}/\omega_b=1.294$ to the saddle-node bifurcation at
which the S state selected by the front is created (see
Fig.~\ref{Bifdiagker}). For mortalities above MI
($\omega_{d0}/\omega_b> 1.345$) the pushed front continues to
be selected over the pulled front that now exists, until the
speed of the latter exceeds that of the pushed front;
thereafter the pulled front prevails
(Fig.~\ref{velocityS-P}(a)). An example of this pulled front
advancing into the P state is shown in Fig.~\ref{frontS-P}. The
transition from pushed to pulled takes place around
$\omega_{d0,c}/\omega_b \approx 1.375$ where an abrupt change
in the dependence of the front speed on the mortality is
clearly visible. This change in the behavior of the front speed
is associated with a similar change in the wavenumber of the S
state deposited behind the front and the wavenumber measured at
the leading edge (Fig.~\ref{velocityS-P}(b)). The pulled S-P
front that prevails at sufficiently high mortalities remains
stable until the saddle-node bifurcation $SN$ of the P state;
its speed is well predicted by the linear marginal stability
calculation throughout this range as is the wavenumber measured
at the leading edge. Similar results have been found in other
systems~\cite{Hari,Archer}.


\subsection{Fronts in version II of the model}

In this section we study briefly the same fronts as in Section
\ref{results_modelI} but for version II of the model. The
simulations are again done with a pseudospectral method,
employing $\Delta t=0.167$ $years$, $\Delta x=0.025$ $m$,
$N=1024$ and periodic boundary conditions. Initial fronts are
formed by connecting smoothly the two desired spatially
extended states. The following figures summarize the results.
As before, we take $\omega_{d0}$ as the control parameter, with
the other parameters as in Fig.~\ref{Bifdiagsimp}.

\begin{figure}
  \includegraphics[width=\columnwidth]{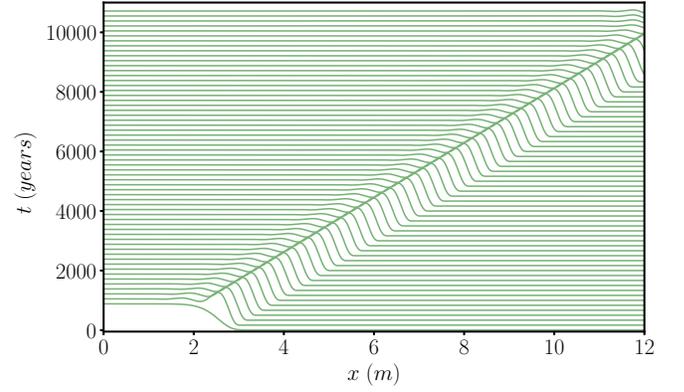}
\caption{Space-time representation of a pushed P-U front in version II of
the model, in which a stable
P state invades a stable U state with speed $v\approx$ $1.091 \cdot 10^{-3}$ $m/year$.
Here $\omega_{d0}=0.069$ $year^{-1}$ ($\omega_{d0}/\omega_b=1.15$) and the remaining parameters are as in
Fig.~\ref{Bifdiagsimp}.
}
\label{frontP-U-simp}
\end{figure}

Figure~\ref{frontP-U-simp} shows an example of a P-U front at
$\omega_{d0}/\omega_{b}=1.15$, i.e., a pushed front connecting
the two homogeneous states P and U, which are both stable at this
mortality value. The figure shows
that P invades U. For the parameter values used the front has
a constant but nonmonotonic profile and travels at constant
speed $v\approx1.091\cdot10^{-3}$ $m/year$.

\begin{figure}
  \includegraphics[width=\columnwidth]{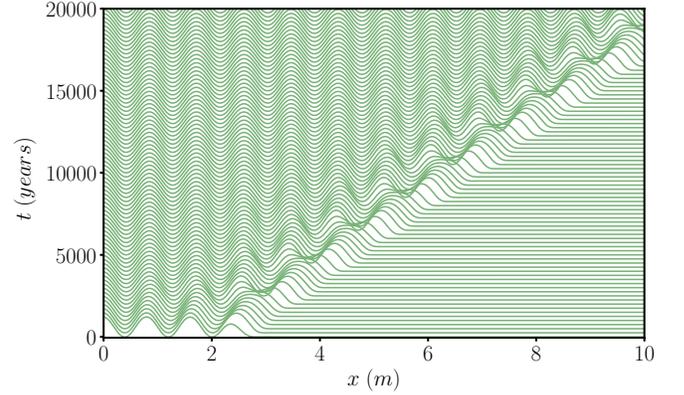}
\caption{Space-time representation of an S-U front in version II of the model.
The front travels with speed $v\approx$ $4.330\cdot 10^{-4}$ $m/year$ in the direction of S invading U.
Here $\omega_{d0}=0.075$ $year^{-1}$ ($\omega_{d0}/\omega_b=1.25$)
and the remaining parameters are as in Fig.~\ref{Bifdiagsimp}.
}
\label{frontS-U-simp}
\end{figure}

\begin{figure}
  \includegraphics[width=\columnwidth]{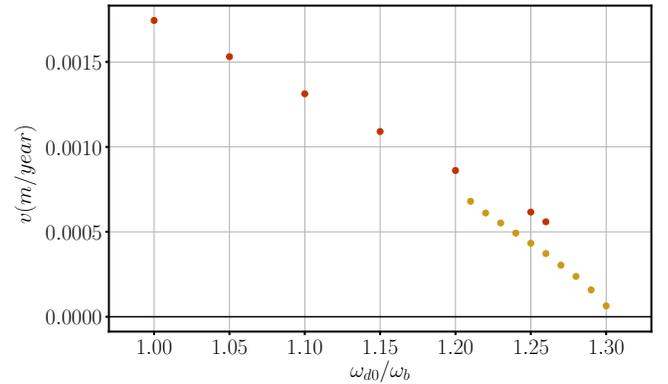}
  \caption{Speed $v$ of P-U (red symbols) and S-U (yellow symbols) fronts in
    version II of the model as functions of $\omega_{d0}/\omega_{b}$. Parameters are
    as in Fig.~\ref{Bifdiagsimp}.
  }
  \label{velocityP-U-simp}
\end{figure}

Figure~\ref{frontS-U-simp} shows a space-time representation of
a S-U front at $\omega_{d0}/\omega_{b}=1.25$, i.e., a front
connecting the stable stripe state S to the stable bare ground
state U. This is a pushed front whereby the stripe state
colonizes bare ground via a time-dependent precursor that
evolves into a stationary stripe pattern. For our parameter
values the invasion speed $v\approx4.330\cdot 10^{-4}$
$m/year$.

Figure~\ref{velocityP-U-simp} shows the speed of S-U fronts as
a function of $\omega_{d0}/\omega_{b}$ and compares it with the
speed of P-U fronts. We see that for fixed parameter values the
latter travel faster, an effect we attribute to the absence of
pinning.



\begin{figure}
  \includegraphics[width=\columnwidth]{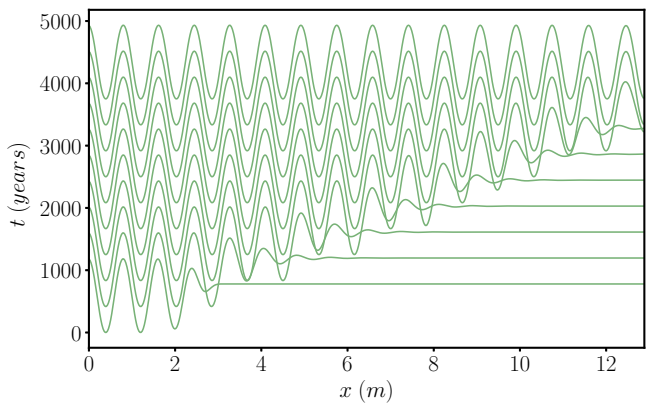}
\caption{Space-time representation of an S-P front in version II of the model.
The front travels with speed $v\approx$ $4.419\cdot 10^{-3}$ $m/year$.
Here $\omega_{d0}=0.078$ $year^{-1}$ ($\omega_{d0}/\omega_b=1.3$)
and the remaining parameters are as in Fig.~\ref{Bifdiagsimp}.
}
\label{frontS-P-simp}
\end{figure}

Figure~\ref{frontS-P-simp} shows a space-time representation of
the third type of front, an S-P front connecting the stripe
state S to the homogeneous populated state P at
$\omega_{d0}/\omega_{b}=1.3$. Both states are stable so this is
a pushed front. In contrast to version I of the model here the front invasion
does not proceed with a clearly defined temporal period between
successive nucleations of new stripes, which leads to a
non-constant velocity. We are unsure of the reason for this.
Nevertheless a mean invasion speed can be estimated and we find
that $v\approx4.419\cdot 10^{-3}$ $m/year$.

\begin{figure}
\includegraphics[width=\columnwidth]{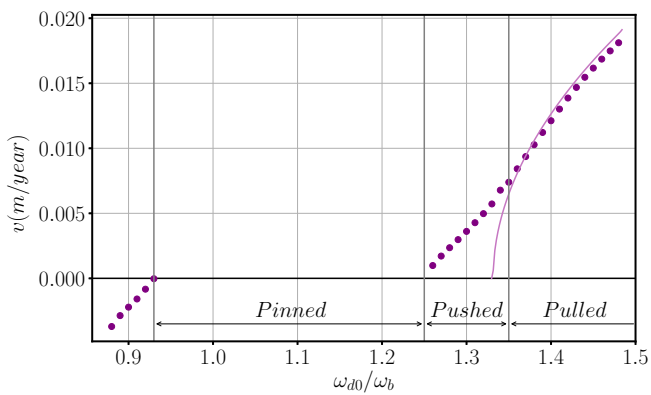}
\caption{Speed $v$ of an S-P front in version II of the model as a function of
$\omega_{d0}/\omega_{b}$, showing the pinning region in which the front is stationary.
The dots correspond to numerical simulations while the solid line represents
the linear marginal stability prediction for a pulled front. Parameters are as in Fig.~\ref{Bifdiagsimp}.
}
\label{velocityS-P-simp}
\end{figure}

Finally, Fig.~\ref{velocityS-P-simp} shows a classic example of
a pinning-depinning transition associated with S-P fronts
\cite{Burke06}. The speed $v$ of the front decreases as one
approaches the edge of the pinning region containing stationary
spatially localized structures; sufficiently close to the edge
the speed is expected to vary as the square root of the
distance from the edge. Within the pinning region the front is
self-pinned, i.e., it is pinned to the pattern state behind it,
and in the depinned regime ($|v|>0$) the front is {\it pushed}
until it is superseded at $\omega_{d0}/\omega_b\approx 1.355$
by the {\it pulled} front created in the MI.

\section*{Conclusions}
\label{S4}

We have explored two versions of a simplified model for clonal plant growth
\cite{ruiz2019ABDaprox}, motivated by undersea patterns
observed in {\it Posidonia oceanica} meadows
\cite{ruiz2017fairy}. The first version takes into account
nonlocal competition and facilitation through appropriately
formulated, albeit phenomenological kernels. The second
simplifies these kernels via a gradient expansion and, after
truncation, leads to a nonlinear but local evolution equation.
In both cases we have taken the mortality parameter
$\omega_{d0}$ as the bifurcation parameter and explored the
behavior of each version as the mortality varies. In both cases
we have made every effort to employ realistic values of the
remaining parameters.

The key findings of our work are:

(i) There is a qualitative agreement between the nonlocal and
local models in that both exhibit the same sequence of
transitions between the three spatially extended states, the
populated state P, the unpopulated state U and the pattern
state S, as $\omega_{d0}$ varies. Nevertheless, substantial
quantitative differences are seen. The nonlocal version is believed
to provide more accurate predictions for the real vegetation
dynamics, whereas the local approximation, because of its
simpler structure, can be used as a qualitative tool to
understand the transitions between different regimes.

(ii) In addition to spatially extended states both systems also
exhibit two types of spatially localized structures, one
resembling holes in the homogeneously populated state and the
other resembling vegetation patches on bare ground, i.e.,
embedded in the U state. These states are organized within
distinct bifurcation structures of snaking type, qualitatively
similar to those arising in related local
\cite{ParraRivas2020formation} and nonlocal
\cite{Cisternas2020gapped} vegetation models.

(iii) Both systems exhibit a variety of fronts connecting
extended states, and these may be either pushed or pulled. In
the former case the speed of the front is determined by
nonlinear processes while in the latter the front
speed can be computed from a marginal stability criterion as
described in~\cite{VanSaarloos2003}. In many systems, pulled
fronts with marginal stability velocities are good descriptions
of fronts describing stable states invading unstable states.
Here we have found situations in which this is the case, but
also cases in which pushed fronts prevail. The characteristic
front speeds are in all cases very slow, of the order of
centimeters per year, a result that is consistent with the
observed slow evolution of {\it Posidonia oceanica} meadows,
the case to which model parameters were fitted.

We emphasize that the results obtained here for the
one-dimensional case are also relevant to the spreading of
vegetation in two dimensions, since stable localized structures
in one dimension block front propagation in two dimensions.
However, our understanding of the dynamics of vegetation fronts
based on the results of a one-dimensional analysis is
necessarily limited since transverse instabilities, if present,
may change both the front profile and its speed. Moreover, the
presence of multiple patterns, their different orientations
with respect to the front, and their different stability ranges
make the analysis of fronts between patterned states in two
dimensions much more challenging.

The spatial period-doubling we observe at small amplitude near
the transcritical bifurcation of the U state appears to be
characteristic of many vegetation models. In the present case
it takes place via peak-splitting as the mortality parameter
$\omega_{d0}$ increases, a process that occurs in related
systems as well~\cite{Yochelis08,Zelnik18}. This process
requires that near their termination the peaks that result
adjust their mutual position to generate a periodic state,
since only periodic states can terminate in a Turing
bifurcation. Other systems exhibit spatial period division
organized within a foliated snaking structure which does not
require the localized structures to adjust their location
\cite{Lloyd2013,Verschueren2017,Zelnik17,Parra2018a,Verschueren2019}.
Related wavelength division is found in other systems~\cite{Parra2018}.
The fact that the region of stability of
periodically spaced vegetation patches appears to extend all
the way down to zero wavenumber (in an infinite domain) allows
sensitive wavelength adaptation when parameters are varied~\cite{Siteur2014}.

\acknowledgements
DRR, LM, EHG and DG acknowledge financial
support from FEDER/Ministerio de Ciencia, Innovaci\'{o}n y
Universidades - Agencia Estatal de Investigaci\'{o}n through
the SuMaEco project (RTI2018-095441-B-C22) and the Mar\'{\i}a
de Maeztu Program for Units of Excellence in R\&D (No.
MDM-2017-0711). D.R.-R. also acknowledges the fellowship No.
BES-2016- 076264 under the FPI program of MINECO, Spain. The
work of EK was supported in part by the National Science
Foundation under grant DMS-1908891, and by a visiting position
at IFISC funded by the University of the Balearic Islands.

\appendix*
\section{Marginal stability predictions for pulled P-U and S-P fronts}

In this Appendix we give details of the marginal stability
predictions for the velocity, and real and imaginary parts of
the wavenumber of pulled P-U and S-P fronts in the two versions
of the model.

\subsection{Version I}
We use the notation $A^*\equiv an^*e^{-an^*}$,
$e_{\kappa}\equiv e^{-(k_r^2-k_i^2)\sigma_\kappa^2/2}$ and
$e_{0}\equiv e^{-(k_r^2-k_i^2)\sigma_0^2/2}$, where $n^*$ is
the constant density of a populated state P given by the
solution of the equation $\omega_b-\omega_d(n^*)=0$. The
dispersion relation obtained from the linearization of version
I of the model around the state $n=n^*$ can be used to write
the condition $vk_i=Re[\lambda(k)]$ in the form
\begin{eqnarray}
  vk_i &=& - 2bn^{*2} -(d_0+d_1n^*)(k_r^2-k_i^2)\nonumber\\
  &&- A^*\bigg(\kappa e_{\kappa}\cos(k_r k_i \sigma_\kappa^2) - \omega_{d0} e_{0}\cos(k_r k_i \sigma_0^2)\bigg).\nonumber\\
\end{eqnarray}
Similarly the condition $Re\left[\frac{d\lambda (k)}{dk}\right]=0$ becomes
\begin{eqnarray}
  0 &=& \bigg( -(d_0+d_1n^*)\nonumber\\
  &+& A^*\bigg(\kappa e_{\kappa}\cos(k_r k_i \sigma_\kappa^2)\frac{\sigma_\kappa^2}{2} - \omega_{d0} e_{0}\cos(k_r k_i \sigma_0^2)\frac{\sigma_0^2}{2}\bigg)\bigg)2k_r \nonumber\\
  &+& A^*\bigg(\kappa e_{\kappa}\sin(k_r k_i \sigma_\kappa^2)\frac{\sigma_\kappa^2}{2} - \omega_{d0} e_{0}\sin(k_r k_i \sigma_0^2)\frac{\sigma_0^2}{2}\bigg)2k_i,\nonumber\\
\end{eqnarray}
while condition $v = -Im\left[\frac{d\lambda (k)}{dk}\right]$ takes the form
\begin{eqnarray}
  v &=& \bigg( +(d_0+d_1n^*)\nonumber\\
  &-& A^*\bigg(\kappa e_{\kappa}\cos(k_r k_i \sigma_\kappa^2)\frac{\sigma_\kappa^2}{2} - \omega_{d0} e_{0}\cos(k_r k_i \sigma_0^2)\frac{\sigma_0^2}{2}\bigg)\bigg)2k_i \nonumber\\
  &+& A^*\bigg(\kappa e_{\kappa}\sin(k_r k_i \sigma_\kappa^2)\frac{\sigma_\kappa^2}{2} - \omega_{d0} e_{0}\sin(k_r k_i \sigma_0^2)\frac{\sigma_0^2}{2}\bigg)2k_r.\nonumber\\
\end{eqnarray}
These three equations are solved numerically for the unknowns
$v$, $k_r$ and $k_i$ characterizing the speed and leading edge
profile of a pulled front, specifically a pulled S-P front as
illustrated in Fig.~\ref{velocityS-P} (solid line). The same
procedure around $n^*=0$ leads to the analytical solution
$v=2\sqrt{d_0(\omega_b-\omega_{d0})}$, $k_r=0$ and
$k_i=\sqrt{(\omega_b-\omega_{d0})/d_0}$ for a pulled P-U front,
as illustrated in Fig.~\ref{velocityP-U} (solid line).

\subsection{Version II}
Following the same procedure for version II the dispersion relation
obtained from the linearization around $n=n^*$ can be written as
\begin{equation}
  \lambda = p_1 + p_2k^2 + p_3k^4,
\end{equation}
where $p_1 = \omega_b-\omega_{d0} - 2a'(\kappa-\omega_b)n^* -3bn^{*2}$, $p_2 = -((\alpha + d_1)n^* + d_0)$, and $p_3 = \beta n^*$. Thus, the condition $vk_i=Re[\lambda(k)]$ can be written in the form
\begin{equation}
  vk_i = p_1 + p_2(k_r^2-k_i^2) + p_3(k_r^4 + k_i^4 - 6k_r^2k_i^2).
\end{equation}
The second condition $Re\left[\frac{d\lambda (k)}{dk}\right]=0$ becomes
\begin{equation}
  0 = 2k_r( p_2 + 2p_3(k_r^2 - k_i^2) - 4p_3k_i^2),
\end{equation}
and the last condition $v = -Im\left[\frac{d\lambda (k)}{dk}\right]$ takes the form
\begin{equation}
  v = -2k_i( p_2 + 2p_3(k_r^2 - k_i^2) + 4p_3k_i^2).
\end{equation}
The velocity of a front between stripes and the homogeneous solution can be computed in the case where $k_r\neq 0$, where
\begin{equation}
 k_r = \pm \sqrt{ \frac{-3p_2\pm \sqrt{7p_2^2-24 p_1p_3}}{8p_3}}.
\end{equation}
The imaginary part $k_i$ can be computed using the expression
\begin{equation}
 k_i = \pm \sqrt{\frac{p_2 +2 p_3 k_r^2}{6 p_3}}.
\end{equation}
In terms of these quantities the speed $v$ is given by
\begin{equation}
 v = -8 k_i p_3 (k_r^2+k_i^2).
\end{equation}
The last expression is used to compute the velocity of a pulled
S-P front, which is represented in Fig.~\ref{velocityS-P-simp}.
The velocity of a P-U front can be obtained from a
linearization around $n^*=0$, which leads to the same
expressions as in version I of the model.



\bibliography{Ref3}


\end{document}